\documentclass[useAMS,usenatbib,twocolumn]{mn2e}

\usepackage{varioref,indentfirst}
\usepackage{graphicx}

\title[Star formation and environment]{Star formation and the
environment of nearby field galaxies}

\author[A. Mateus Jr. \& L. Sodr\'e Jr.]{Ab{{\'\i}}lio Mateus Jr.$^{1}$
\thanks{E-mail: abilio@astro.iag.usp.br; laerte@astro.iag.usp.br} 
and Laerte Sodr\'e Jr.$^{1}$
\footnotemark[1]\\
$^{1}$Departamento de Astronomia, IAG-USP, Rua do Mat\~ao 1226, 05508-090, 
S\~ao Paulo, Brazil}

\begin{document}

\date{\today}

\pagerange{\pageref{firstpage}--\pageref{lastpage}} \pubyear{2004}

\maketitle

\label{firstpage}

\begin{abstract}
We investigate the environmental dependence of galaxies with star 
formation from a volume-limited sample of $4782$ nearby field galaxy 
spectra extracted from the 2dF Galaxy Redshift Survey final data release. 
The environment is characterized by the local spatial density 
of galaxies, estimated from the distance to the 5th nearest neighbour.
Extensive simulations have been made to estimate correction factors
for the local density due to sample incompleteness.
We discriminate the galaxies in distinct 
spectral classes -- passive, star-forming, and short starburst galaxies -- 
by the use of the equivalent widths of [{\sc O~ii}]$\lambda$3727 
and H$\delta$. 
The frequency of galaxies of different classes are then evaluated 
as a function of the environment. We show that the fraction of
star-forming galaxies decreases with increasing density, whereas passive 
galaxies present the opposite behaviour. 
The fraction of short starburst galaxies -- that suffered a starburst
at $\sim$ 200 Myr ago -- do not present strong environmental dependence. 
The fraction of this class of galaxies is also approximately constant
with galaxy luminosity, except for the faintest bins in the sample, where
their fraction seems to increase. We find that the star-formation properties
are affected in all range of densities present in our sample (that excludes
clusters), what supports the idea that star-formation in galaxies is affected 
by the environment everywhere. We suggest that
mechanisms like tidal interactions, which act in all environments, do play a 
relevant role on the star-formation in galaxies.

\end{abstract}


\section{INTRODUCTION}

It has become clear along the last decades that the environment a
galaxy inhabits has a profound impact on its star-formation properties.
Indeed, the study of the variations of star formation in galaxies 
as a function of their environment has a great importance for 
extragalactic astronomy, since it makes possible to 
understand some properties of the formation and evolution of 
galaxies in the Universe, helping to discriminate between `nature'
and `nurture' effects.

It is well known that at low redshifts star-forming 
galaxies constitute a small 
population in high-density regions, especially in galaxy clusters.
\citet{osterbrock60} observed that the frequency of emission-line 
galaxies
is lower in clusters than in the field. Later, this trend was
verified in a quantitative way by many works 
\citep[][and others]{gisler78,dress85,abraham96,ba99,po99,loveday99,
ellingson01}.
Furthermore, it is also known the trend for {\sc H~i} deficiency 
among cluster spirals
galaxies \citep{giovanelli85,solanes96,bravo00,goto03}, 
that could explain
the interruption of star formation in terms of the depletion of the
galaxy gaseous content, either by its consumption or suppression 
\citep{gisler78,ken83,kenny89}. On the other side, some works show 
that spirals in clusters may have a star formation rate similar or 
even larger than
field spirals \citep{gavazzi85,moss93,gavazzi98,biviano97}, although, as
suggested by Gavazzi \& Jaffe, the observed increment in the 
star-formation rate of these galaxies may be a transient phenomenon, 
which occurs while the galaxy suffers the suppression of its 
gaseous content.
In low-density environments, tidal interactions are the most relevant process.
Ordinary star-forming galaxies have an external gaseous reservoir that
is critically important to the continuous gas supply of the galaxy disk.
Numerical simulations \citep{bekki01a} show that this reservoir is fragile 
and, consequently, more susceptible to tidal effects. 

Effects of the environment on the star-formation 
in galaxies, then, must involve at least two types of processes: 
i) those that decrease the gaseous content and, therefore, reduce 
the potential of star formation in galaxies, and
ii) processes that trigger bursts of star formation. 
Among the first class of process there are: interactions between 
the intragalactic 
and intergalactic medium, including gas removal and evaporation 
\citep{gunn72,fujita99}; tidal interactions, that remove the gas of 
the disk of spiral galaxies \citep{byrd90}; suppression of the accretion 
of gas-rich materials in the neighbourhood of the galaxy \citep{LTC,bekki01a}. 
The second type of processes include: gas compression by ram-pressure, 
that induces star formation \citep{dressler83,bothun86,vollmer01}; 
fusion with other systems \citep{barnes91,lavery94,bekki01b}; 
tidal interactions \citep{moss00}. Thus, process like tidal
interactions may trigger star-formation as well as may contribute to
its end.

Recent galaxy surveys are allowing to investigate the relation between
star formation and environment with large samples of data collected
in a uniform way. This includes the Las Campanas Redshift Survey 
\citep{hashi98},  the 15R-North Galaxy Redshift Survey \citep{carter}, 
the 2dF Galaxy Redshift Survey
\citep[2dFGRS,][]{lewis02}, and the Sloan Digital Sky Survey
\citep[SDSS,][]{gomez02}. \citet{lewis02} have analyzed the
environmental dependence of star formation rates (SFRs) near galaxy clusters
in the 2dFGRS region, finding that there is a correlation between
SFR and projected density, that disappears below $\sim1$ h$_{70}$ galaxy 
Mpc$^{-2}$ (for galaxies brighter than $M_b=-19$). 
The SDSS data shows a strong correlation between SFR and local projected 
density, with a ``break'' in this relation at the same density found 
near the 2dFGRS clusters; below this threshold, corresponding to a 
clustercentric radius of $\sim$~3 virial radii, the SFR varies slowly
with local projected density.

In this work, {instead of focusing on the SFR,} 
we analyse the environmental dependence of 
the population of star-forming nearby field galaxies 
($z < 0.05$) based on a 
volume-limited sample extracted from the 2dFGRS final data 
release \citep{colless03}.
The star formation is characterized by spectral classes
defined using the equivalent widths of [{\sc O~ii}]$\lambda$3727 and H$\delta$.
The environmental parameter is the local spatial density of
galaxies. The paper is organized as follows. Section 2 presents the 
sample selection, the method adopted for the density estimation, and 
the spectral indices that will be analysed here.
Section 3 describes the 
spectral classification and presents the relation between 
the fraction of distinct galaxy types and environment in 
our sample. Our results are discussed in Section 4. Finally, in Section 5 we 
summarize our conclusions.

\section{The sample and the estimation of local galaxy density}

In this section we describe the sample of galaxies that will be
analysed and present the parameter that will be used to describe a galaxy
environment, the local galaxy density. We also present the spectral indices
that will be used to describe star-formation properties of the galaxies.
Finally, we discuss the selection function of the sample.

\subsection{Limited volume sample}

The data used in this work were extracted from the 2dFGRS spectra
recently available on the final data release
\citep[][see also Colless et al. 2001 for further and detailed 
information about the survey]{colless03}\footnote{The 2dFGRS database 
and full documentation are available on the WWW at \tt{http://www.mso.anu.edu.au/2dFGRS/}}.
The 2dFGRS obtained spectra for 245591 objects, 
mainly galaxies, brighter than an extinction-corrected 
magnitude limit of $b_J=19.45$.
We have constructed a limited volume sample with 
galaxies within the strips located in the Northern 
(NGP; $2.5^{\circ} > \delta > -7.5^{\circ}$, 
$9^{h}50^{m} < \alpha < 14^{h}50^{m}$) and in the Southern Galactic
hemispheres (SGP; $-22.5^{\circ} > \delta > -37.5^{\circ}$, 
$21^{h}40^{m} < \alpha < 3^{h}40^{m}$). The sample comprises galaxies
with radial velocities between 600 and 15000 km s$^{-1}$ 
brighter than an extinction corrected absolute magnitude 
$M_{b_{J}}^{lim}=-17.38 + 5\log h$, corresponding to corrected 
apparent magnitudes $b_{J_{lim}} \le 18.50$.
The advantage of this approach is that the radial selection function
is uniform (in the case of complete sampling) and variations 
in the space density of galaxies within the volume are due to clustering 
only \citep{norb02}. Considering only galaxies with spectra with
quality parameter $Q \ge 3$ \citep{colless01}, this 
initial sample contains 8040 galaxies.
As will be shown in Section 3.2, the fraction of star-forming galaxies
increases with the limiting absolute magnitude of the sample, and
hence our results regarding the fraction of spectral classes are
also dependent on this parameter.

Since we are interested in field galaxies, it is necessary to remove
from the sample galaxies that are members of galaxy clusters. 
The clusters within the 2dFGRS region were studied 
by \citet{propris02},
and their cluster catalogue may be considered complete up to
$z = 0.1$, well above the limit of our volume limited sample.
The studies of \citet{gomez02} and \citet{lewis02}
for the SDSS and 2dFGRS, respectively, indicate that a representative
sample of field galaxies cannot be obtained within $\sim$~3 to 4 virial
radius of the cluster core. The virial radius of a cluster can be calculated 
as $R_V = 0.002 ~ \sigma_r ~ h^{-1}$ Mpc, where $\sigma_r$ is the radial
velocity dispersion in units of km s$^{-1}$ \citep{girardi98}, and for 
the clusters studied by \citet{gomez02} its mean value is $\sim~1~h^{-1}$ Mpc.
Consequently, we excluded from the sample (as well as from the simulations
described below) all galaxies inside a sphere of 4~$h^{-1}$~Mpc radius around 
the cluster centers catalogued by \citet{propris02}.
The resulting sample of field galaxies contains $6768$ galaxies.

\begin{figure*}
\centerline{\includegraphics[width=175mm]{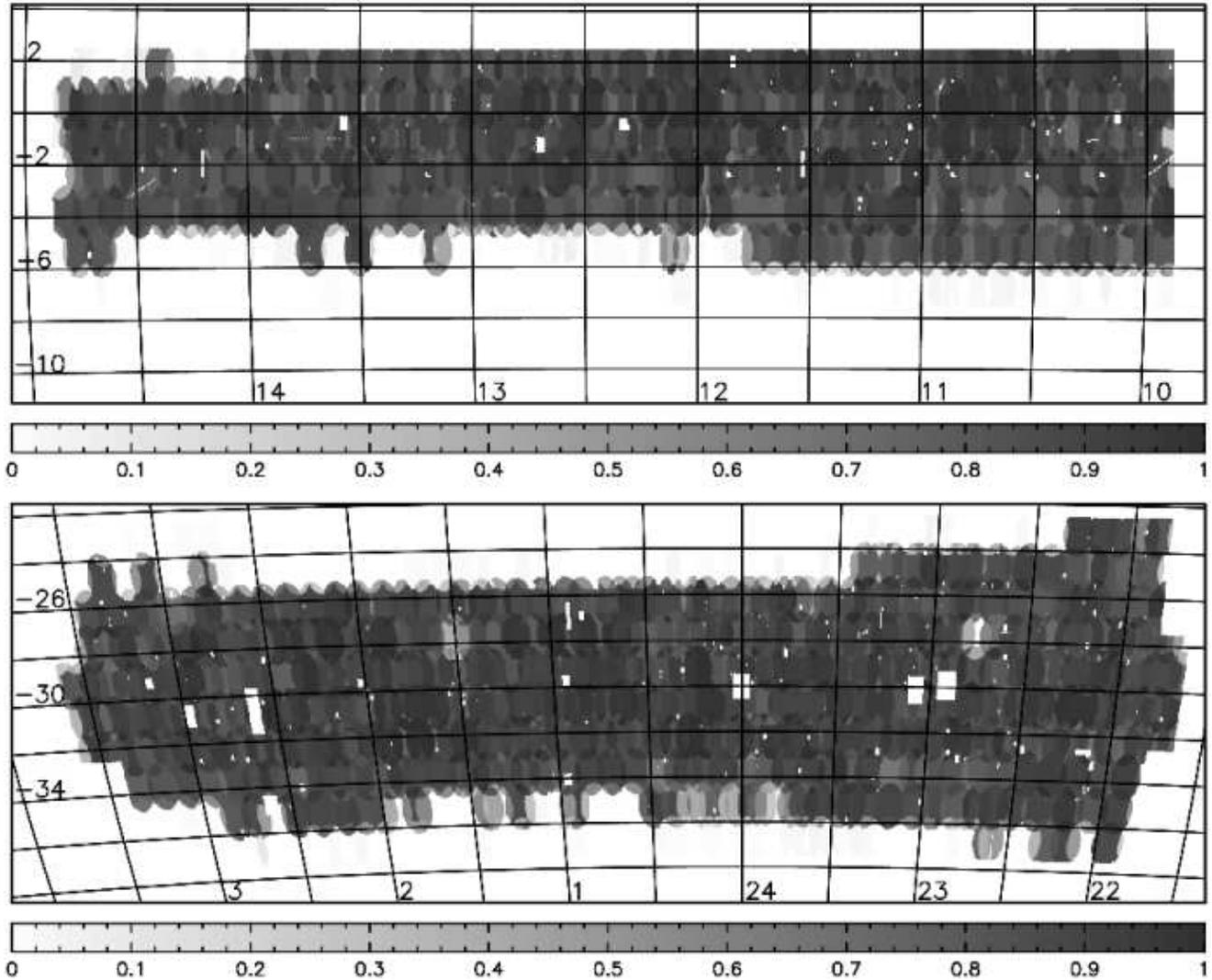}}
\caption{Redshift completeness as a function of position for
the simulated galaxies. The top panel is for the NGP strip and the  
bottom panel is for the SGP strip. The horizontal axis represents the
right ascension (in hours) and the vertical axis is the declination 
(in degrees). The completeness, between 0 and 1, is represented by different 
levels of intensity, with darker regions representing regions of higher
completeness, following the scale shown below the strips.}
\label{1}
\end{figure*}

\subsection{Simulations} 

Despite the fact that the 2dFGRS final data release 
has a high completeness, this is not true for 
some regions covered by the survey, and 
we need to take into account the possibility of survey 
incompleteness in the determination of the environment associated to 
each object in our sample.
With this aim, we generated mock catalogues to simulate
the distribution of galaxies in our selected volume and the distribution 
of the objects selected by the 2dFGRS. Firstly, we computed the 
mean galaxy number density brighter than the adopted luminosity limit: 
\begin{equation}\label{LF}
\bar{\rho}(>L_{min})=\int^{\infty}_{L_{min}} \Phi(L) dL
\end{equation}
where $\Phi(L)$ is the Schechter luminosity function with parameters
(in $b_{J}$ band)  $M^{*}=-19.66 + 5 \log h $, $\alpha=-1.21$ and
$\Phi^{*}=1.66\times10^{-2}\,h^3$Mpc$^{-3}$ \citep{norb01}, that
are appropriate for the 2dFGRS. Then, 
for the NGP and SGP regions, we evaluated the mean galaxy 
number in each region, $\bar{N}$, that would be expected for a uniform galaxy 
distribution in each volume. After, for each 
simulation, we assumed a Gaussian distribution (with mean $\bar{N}$
and dispersion  ${\bar{N}}^{1/2}$) to determine the actual number 
of galaxies, $N$, included in each volume. Finally, for 
the $N$ galaxies in each region we have randomly selected 
a position ($\alpha$, $\delta$), a luminosity (based on the 
Schechter luminosity function) and a redshift, assuming
a uniform distribution within each volume.

At this point we need to verify how the incompleteness of the survey 
affects the simulated catalogues. \citet{colless01} present a 
description of the magnitude and completeness masks that were constructed 
for this purpose. The redshift completeness is given by a parameter
called $R$, which depends on the position at the strips covered by the 
survey and is the ratio between the observed 
number of objects and the total number of objects in the parent
catalogue at that position. Moreover, the
masks supply a magnitude limit and a $\mu$ parameter that also depends
on the position. The redshift completeness is magnitude dependent 
and, as shown by \citet{norb01}, it can be written as
\begin{equation}
c_{z}(b_{J},\mu_{i})=0.99\;[1-\exp\,(b_{J}-\mu_{i})].
\end{equation}
Thus, in a second stage of the simulations, we have 
applied the completeness masks to our 
simulated data to select a sample
of simulated galaxies with spectra within each region.
In this way, each simulation results, for each region, in two volume-limited 
samples, one that represents the `parent' galaxy distribution in the
range of redshifts and magnitudes considered here, and another simulating the
sub-sample of galaxies with 2dFGRS spectra.
This approach allows to define a correction factor to the
measured local galaxy density, that will be discussed in the next
section. In Fig. \ref{1} we show the 
completeness map resulting from the simulations for the two 
strips covered by the survey (for comparison, see maps shown in 
Colless et al. 2003).

\subsection{Local spatial density of galaxies}

In recent studies, the environment has been 
characterized either by the local number galaxy density 
\citep[e.g., ][]{hashi98,carter}, or by the 
projected galaxy density \citep[e.g., ][]{lewis02,gomez02}. 

We adopted a non-parametric method to determine the local number density of  
galaxies, based on the $k$th nearest neighbour  density estimator
($k$NN). This method fix a value for $k$ and let the volume  $V$, centred on
a given object and extending to its $k$th nearest neighbour, be a 
random variable. This volume is large in low
density regions and small in high density regions. The $k$NN density 
estimator may be written as \citep{casertano85,fuku}:
\begin{equation}\label{densidade}
\rho=\frac{k-1}{V(r)}
\end{equation}
with $V(r)=4 \pi r^{3}/3$, where $r$ is the distance to $k$th nearest 
neighbour. For our purposes we have used the value $k=5$. We have made tests
for several values of $k$, concluding that 5 is an adequate choice, given
the shape of the survey regions, that does not favour larger values. 
Additionally, we have prevented an incorrect density estimate due to 
border effects by excluding galaxies whose $k$th neighbours 
have projected distances greater than the distance of the 
galaxy to the closest border of the survey region or of the sample volume.

To correct the estimated density for sample incompleteness, we have
generated $800$ pairs of simulated catalogues, each pair
comprising a `parent' and an `observed' sample. This procedure
allowed us to compute a mean local correction factor for the density,
$\mathcal{C}$, given by
\begin{equation}
\mathcal{C}=<\frac{\rho_p}{\rho_o}>,
\end{equation}
where $\rho_p$ is the number density associated with the `parent' sample and
$\rho_o$ is the same for the `observed' sample. 

\begin{figure}
\centerline{\includegraphics[width=84mm]{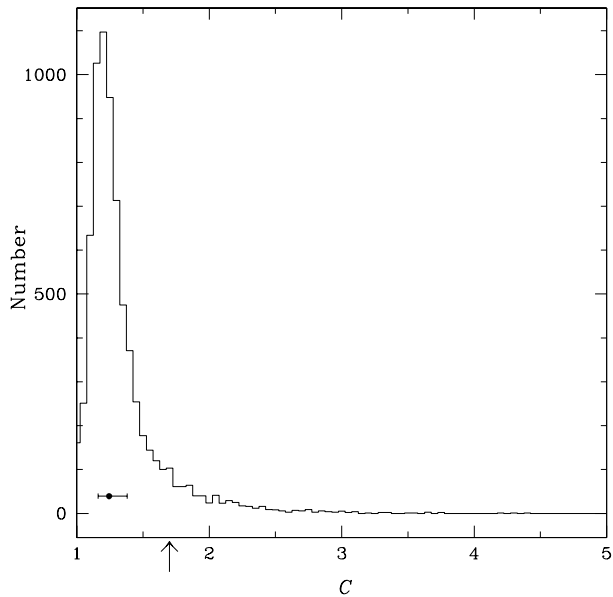}}
\caption{Distribution of the local galaxy density correction factor 
($\mathcal{C}$). The filled circle represents the median
value and the error bars are the respective quartiles of the distribution
of $\mathcal{C}$ values. The arrow  indicates the upper limit in 
$\mathcal{C}$ adopted to select the sample.}
\label{2}
\end{figure}

\begin{figure*}
\centerline{\includegraphics[width=175mm]{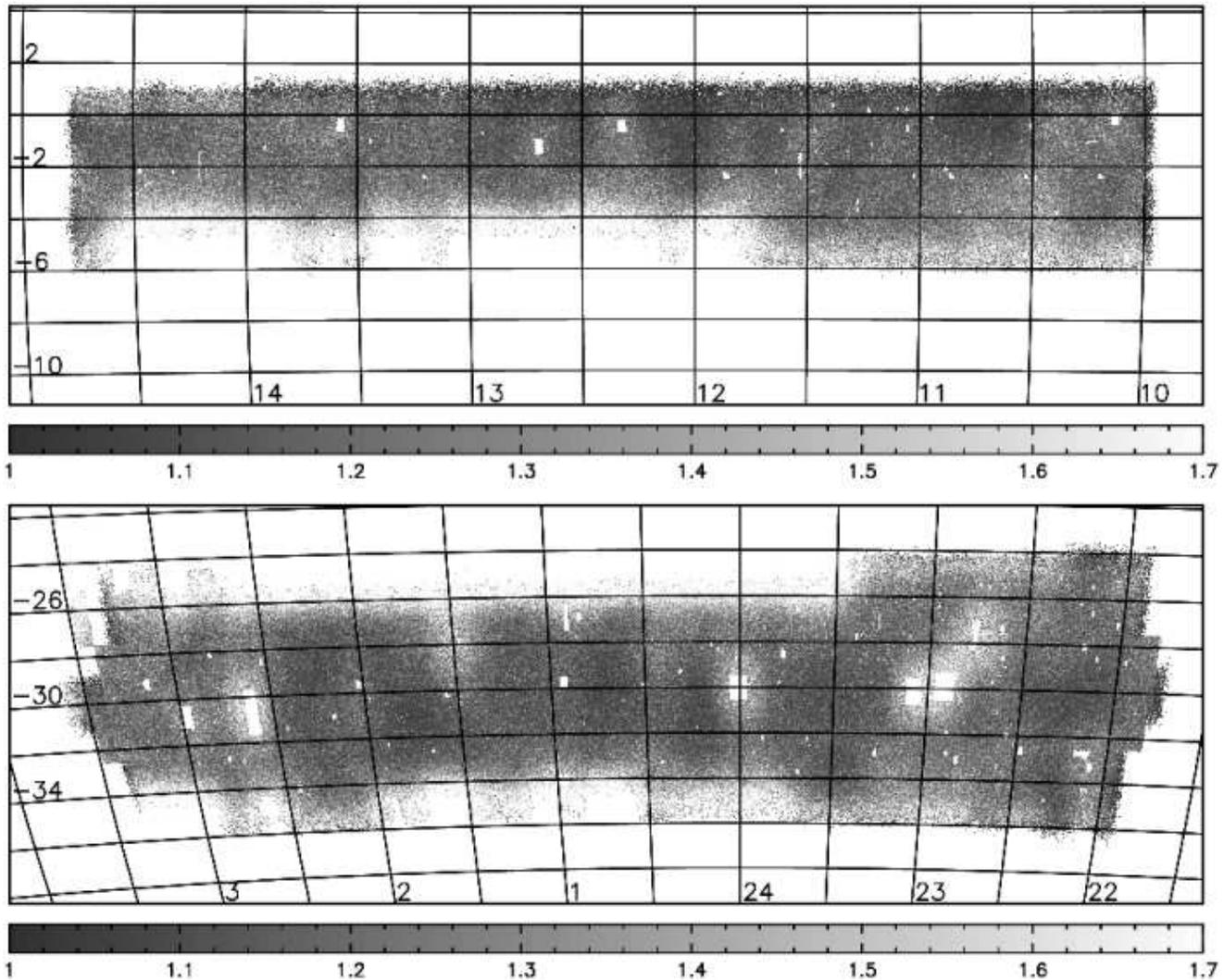}}
\caption{Correction factor, $\mathcal{C}$, as a function of position 
for the selected regions. The top panel is for the NGP strip and the
bottom panel is for the SGP strip. The horizontal axis represents the
right ascension (in hours) and the vertical axis is the declination 
(in degrees).The values of $\mathcal{C}$ are represented by different 
levels of intensity, with darker regions representing regions of lower
corrections, accordingly to the scale shown below the strips.}
\label{3}
\end{figure*}

The local values of $\mathcal{C}$ 
were then used to correct the density associated to 
each galaxy in our initial volume-limited sample.
Fig. \ref{2} shows the distribution of $\mathcal{C}$
for this sample. The filled circle represents the 
median value of the distribution and the error bars are the 
respective quartiles. Note that, by definition, $\mathcal{C} \ge 1$.
To avoid uncertainties introduced in the density by large corrections,
we have restricted the sample to those objects located in regions 
where $1 \le \mathcal{C} \le 1.7$; in Fig. \ref{2}
this limit is indicated by an arrow. 

Since we are neglecting peculiar velocities in distance estimates, 
galaxies in dense environments, where the velocity dispersion is high, 
will probably have their local densities underestimated.
However, since we have removed from the sample 
galaxies within $\sim4$ virial radius around the clusters in the 2dFGRS
survey area (c.f. Sect. 2.1), our results should not be 
strongly affected by peculiar velocities.

At this point, the sample volume is $\sim 14400\,h^{-3}\,$Mpc$^3$ 
and contains $5463$ galaxies.
In Fig. \ref{3} we present the correction factor as a
function of position for the two strips covered by the survey,
where we only show the regions comprising the selected sample that
follows the constraints on $\mathcal{C}$.

\subsection{Measurements of spectral indices}\label{spectral}

For each galaxy in our sample we measured the total (i.e, emission plus
absorption)  equivalent widths 
(EWs) of [{\sc O~ii}]$\lambda$3727, H$\delta$, H$\beta$, 
[{\sc O~iii}]$\lambda$5007, 
[{\sc N~ii}]$\lambda$6548, H$\alpha$ and [{\sc N~ii}]$\lambda$6583 
directly from the 2dFGRS spectra; hereafter these quantities will be 
called `spectral indices'. We adopt here positive values for emission line
EWs, and negative for absorption EWs.
The indices were computed by fitting the continuum defined in two regions 
around the line (blue and red continuum) and measuring the line flux 
normalized relative to this continuum. 
The EW errors were computed following the prescription of \citet{cid}, that
takes into account the noise within the line window and the uncertainty in the
positioning of the continuum.

In Section 3 we will make a spectral classification based on the 
[{\sc O~ii}]$\lambda$3727 and H$\delta$ EWs. This pair of indices is
convenient because it can be measured even at high redshifts and, thus, it
can be used directly to probe evolutionary effects in samples of
more distant galaxies. 
The median values of these indices are very similar to those obtained by 
\citet{ba99} for a sample of field galaxies extracted from the CNOC1 sample;
our EW[{\sc O~ii}] has also mean and median values comparable to those
obtained by \citet{gomez02} for a SDSS sub-sample.
The median errors (with the quartiles of the error distribution) are $3.3^{+1.7}_{-0.9}$ 
and $1.8^{+1.0}_{-0.6}$~{\AA} for
[{\sc O~ii}]$\lambda$3727 and H$\delta$, respectively. The error
distribution has a long tail towards large errors and, thus,
we excluded from the analysis $456$ 
objects with EW uncertainties larger than $10$~{\AA} for EW([{\sc O~ii}]) 
and $6$~{\AA} for EW(H$\delta$). 
Additionally, following \citet{lewis02}, we excluded $225$ galaxies 
with EW(H$\alpha$)~$> 10$~{\AA} and 
EW([{\sc N~ii}]$\lambda$6583)~$> 0.55\,$EW(H$\alpha$).
These objects are classified as active galactic nuclei (AGNs) and
have been removed from the sample because 
they have a significant non-thermal component \citep{vo87},
contrarily to  typical star-forming {\sc H~ii} regions.

A point that deserves mention here is the bias that may be introduced in the
analysis due to the use of small fibers to measure the galaxy spectra. 
This effect, known as aperture bias, is discussed in detail by 
\citet{kochanek}, who demonstrated that it can lead to an 
underestimate of EW values, and consequently, to
an overestimate of the fraction of early-type galaxies in a survey.
\citet{mad02} discuss the presence of this effect in 2dFGRS spectra,
concluding that it does not introduce any significant bias in the 
fractions of galaxies distinguished accordingly to their spectral types.
We will return to this point in Section 4.

\subsection{The selected sample}\label{selecao}

After the exclusion of the objects with high uncertainties in the
spectral indices and those identified as 
AGNs, the final sample comprises $4782$ galaxies, whose properties
will be analysed and discussed in the following sections.

\begin{figure}
\centerline{\includegraphics[width=84mm]{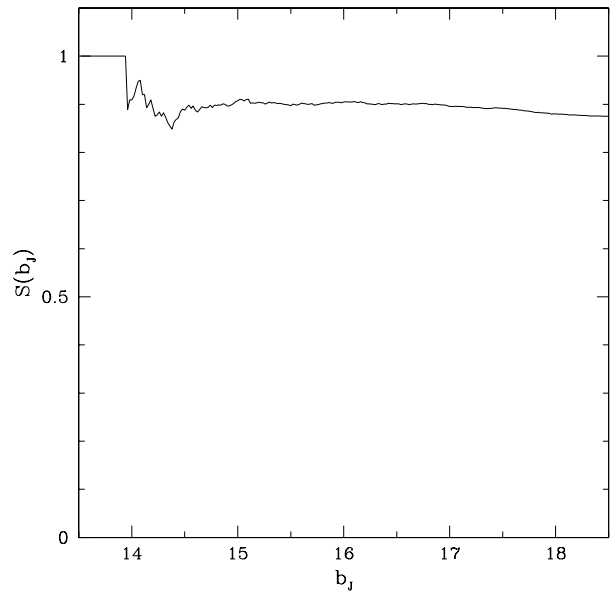}}
\caption{Selection function as a function of apparent magnitude
(b$_{J}$) for the selected sample.}
\label{4}
\end{figure}

\begin{figure}
\centerline{\includegraphics[width=84mm]{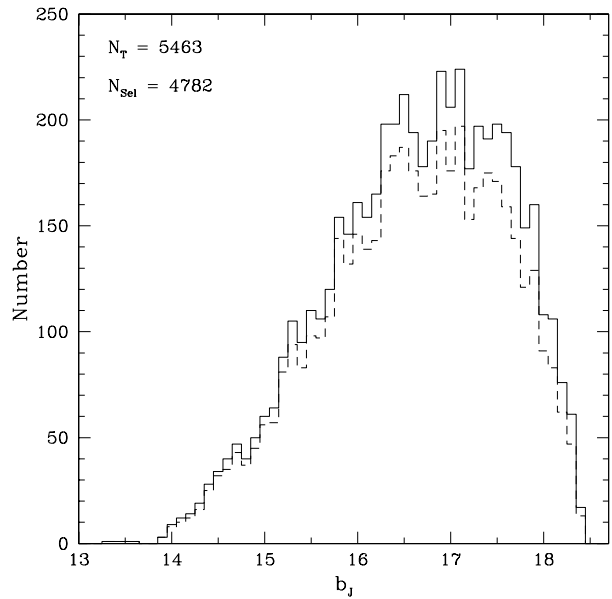}}
\caption{Magnitude distribution for the initial sample 
(with N$_{T} = 5463$ galaxies), shown as a solid line, and for the selected 
galaxies (N$_{Sel} = 4782$), shown as a dashed line.}
\label{5}
\end{figure}

The sample completeness may be described by a selection function
$S(b_{J})$ that we present in Fig. \ref{5}. This 
function is defined as
\begin{equation}
S(b_{J})=\frac{N_{sel}}{N_{T}},
\end{equation}
where, for a given magnitude bin, 
$N_{sel}$ is the number of selected galaxies and
$N_{T}$ is the total number in the volume.
Fig. \ref{4} shows that the selection function remains constant,
and with maximum value, until $b_{J} \sim 14.0$. For
$14.0 \la b_{J} \la 14.5$ the function presents a variable 
behaviour and, for fainter magnitudes, it is approximately constant,
decreasing for $b_{J} \ga 17.0$ and reaching the final value $S(b_{J}) \sim 0.88$.
Fig. \ref{5} shows the distribution of magnitudes for
the selected objects (dashed line) and for the initial sample 
(solid line). This figure suggests that our selection procedures
did not introduce any significant bias in the magnitude distribution
of our final sample.

\section{SPECTRAL CLASSIFICATION AND ANALYSIS}

\subsection{Spectral classification}\label{classification}

Several authors have used EWs in galaxy classification 
\citep[e.g.,][]{hamilton85,po99,ba99,bekki01,po01}. 
One advantage of EWs is that they are not affected by extinction,
although they are sensitive to differences in the extinction of
the regions that produce the emission lines and those producing the 
continuum \citep[the selective extinction; e.g., ][]{stasinska01}.
 Moreover, EWs are relatively 
insensitive to changes in instrumental resolution and they can
be measured in non-flux calibrated spectra, like the 2dFGRS data.

Here we adopt the [{\sc O~ii}] and H$\delta$ EWs to classify 
the galaxies in the sample. These lines can be measured up to large redshifts,
and thus our results may be considered a $z=0$ calibration for
evolutionary studies
of the environmental dependence of the fraction of star forming galaxies.
The EW of the [{\sc O~ii}]$\lambda$3727
doublet is associated with the presence of young, massive
stars, and is a useful tracer of star formation at the blue
side of an optical spectrum. For this reason it has been
adopted in many studies of star formation properties in galaxies.
In fact, works as those of \citet{gallagher89}, \citet{ken92a}, and 
\citet{tresse99} have shown that the [{\sc O~ii}] line presents a
good correlation with primary tracers of star formation, 
like H$\beta$ and H$\alpha$ emission lines. 
The other spectral index we adopted for the spectral classification is 
the EW of the Balmer line H$\delta$. When in emission, this feature is 
produced in objects with high increment in the star formation rate. On the
other hand, strong absorption in H$\delta$ is
associated to galaxies which had an intense starburst
ended about 1--2 Gyr ago \citep{barbaro97}.
Following \citet{ba99}, in this work we will 
classify galaxies in spectral classes accordingly to their position 
in the EW([{\sc O~ii}])--EW(H$\delta$) plane,
which relates an index linked to star formation activity to
another associated with starburst age. Additionally,
we define an object as a star-forming galaxy if it has 
EW([{\sc O~ii}]) $> 5\;${\AA}, that is about $1.5\sigma$ above our 
detection limit. It is also important to note again that we use
positive values to indicate emission lines and negative values
for absorption lines. We adopt here 3 spectral classes \citep[see ][]{ba99}: 
\begin{itemize}
\item {passive galaxies (P: EW([{\sc O~ii}]) $\le$ 5 {\AA}):
galaxies without evidence of significant 
current star formation (in 
general E or S0);}
\item {short starburst galaxies (SSB: EW([{\sc O~ii}]) $>$ 5 {\AA},
EW(H$\delta$)~$>$~0):
galaxies where a large fraction of the light comes from a starburst
that started less than $\sim$ 200 Myr ago;}
\item {ordinary star-forming galaxies (SF: EW([{\sc O~ii}]) $>$ 5 {\AA},
EW(H$\delta$) $<$ 0):
includes most normal spirals and irregulars, that have been forming
stars for several hundred million years.}
\end{itemize}

\begin{figure}
\centerline{\includegraphics[width=84mm]{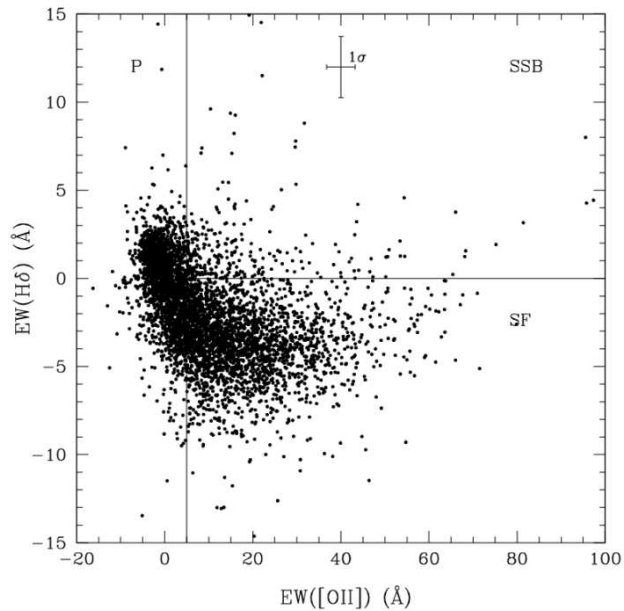}}
\caption{The EW([{\sc O~ii}])--EW(H$\delta$) plane, indicating the
regions that define the three spectral classes (P, SF, SSB) discussed
in this work.}
\label{6}
\end{figure}

Fig. \ref{6} shows the EW([{\sc O~ii}])--EW(H$\delta$) plane with
the distribution of galaxies in the regions of each 
spectral class. The total number of galaxies with star formation is
$N_{TSF}=2750$, of which $N_{SSB}=285$ and $N_{SF}=2465$, and the number of
those without evidence of ongoing star formation is $N_P=2032$. 
Considering only galaxies with star formation, SSBs correspond to  
$\sim 10$\% of the total. This fraction compares very well with 
that obtained by Balogh et al. for a sample of field galaxies 
($\sim 9$\% of star-forming galaxies).
Note that the classification of Balogh et al. also
includes other two classes: K+A e A+em. These classes comprise 
galaxies with strong H$\delta$ absorption 
(EW(H$\delta$)~$<$~$-5$ {\AA}). In our classification, the K+A
galaxies belongs to the P class and the A+em are in the SF class.

It is worth mentioning that the 2dFGRS database provides for each galaxy a
spectral type, derived from a Principal Component Analysis of the spectra
\citep{mad02}. These spectral types are strongly correlated with 
EW(H$\alpha$) and, consequently, they follow the same trends than 
EW(H$\alpha$).

\subsection{Correlations with luminosity}\label{sfluminosity}

\begin{figure}
\centerline{\includegraphics[width=84mm]{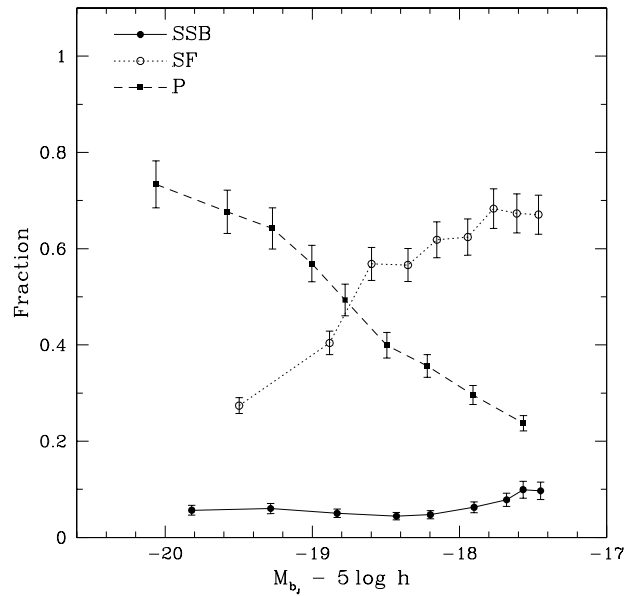}}
\caption{Fraction of galaxies in each spectral class as a function
of the absolute magnitude. For each class, each magnitude 
bin contains about the same number of objects; the
error bars were computed assuming a Poissonian statistics.}
\label{7}
\end{figure}

The fraction of galaxies in each spectral class (relative to the total number 
of objects in the selected sample), is shown in Fig. 
\ref{7} as a function of the absolute magnitude. The fractions of 
SF and P classes show a strong correlation with luminosity:
low-luminosity objects are essentially galaxies that present evidence
of star formation activity, generally late-type spirals and irregulars;
the opposite is seen for brighter objects, which include 
mainly passive galaxies. The population of SSB galaxies present an 
excess of objects with low luminosity, with their fraction increasing 
for $M_{b_J} - 5 \log h \ga -18$; SSB galaxies brighter than this value 
do not show trends with luminosity.

Relations between star formation and luminosity as those shown in Fig. 
\ref{7} have been detected in many surveys. Indeed, the
luminosity function of nearby galaxies, divided accordingly to the
presence or lack of the [{\sc O~ii}] emission line, has been calculated
for the Las Campanas Redshift Survey \citep{lin96} and the
ESO Slice Project \citep{zucca97}, as well as for the
2dFGRS \citep{folkes99,mad02}. These studies have found that 
star-forming galaxies tend to be less luminous than passive galaxies. 
These results are also confirmed by studies of luminosity function of
samples selected by morphological \citep{marzke98} and spectral types 
\citep{bromley98, folkes99}, which have shown that ellipticals and
lenticulars tend to be brighter than late-type spirals and irregulars. 

\begin{figure}
\centerline{\includegraphics[width=84mm]{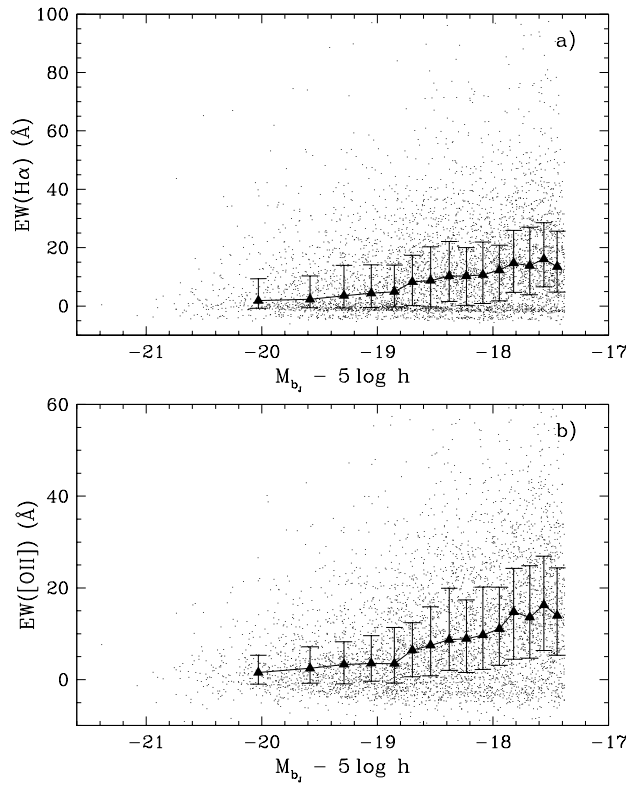}}
\caption{EW(H$\alpha$) and EW([{\sc O~ii}]) as a function of absolute
magnitude. The triangles are median values evaluated in
magnitude bins containing about the same number of objects.
The error bars are the quartiles of the EW distribution within each bin.}
\label{8}
\end{figure}

Fig. \ref{8} shows the luminosity dependence of 
H$\alpha$ and [{\sc O~ii}] EWs. There is a clear trend indicating that
these EWs tend to decrease with increasing luminosity, in a fashion 
analogous to Fig. \ref{7}. The  EW of H$\alpha$ is related
to the ratio between the UV flux emitted by young stars and the flux 
from the old stellar population that produces most of the continuum at the
line wavelength. Thus, a large EW is due either to a large UV flux and/or 
to a small continuum from the old stars. Consequently, galaxies with high
EW(H$\alpha$) form a blue population, while those without 
H$\alpha$ emission are preferentially redder  \citep{kenken83}. 
The same trend is observed in the case of  [{\sc O~ii}]$\lambda$3727.
As suggested by \citet{tresse99}, the ionization sources which are 
responsible for H$\alpha$ emission 
are also related to the [{\sc O~ii}] emission, as well as to other metallic emission 
lines, like those from [{\sc S~ii}], indicating that
low-luminosity star-forming galaxies tend to have large emission line EWs in 
both the blue and the red side of the optical spectrum.

\subsection{Environmental distribution of spectral classes}

\begin{figure}
\centerline{\includegraphics[width=84mm]{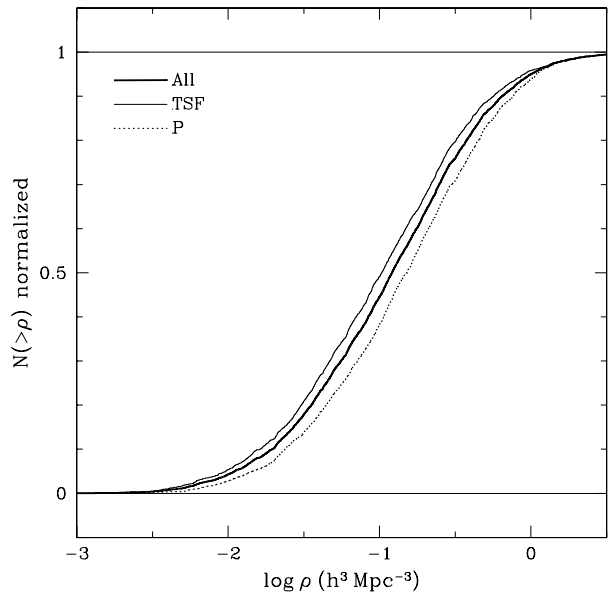}}
\caption{\small Cumulative distribution of the number of galaxies
as a function of the density for all galaxies in the sample (thicker solid line),
for all galaxies with star formation (SF + SSB; solid line), and
for passive galaxies (dotted line).}
\label{9}
\end{figure}

The normalized cumulative distribution of all galaxies in the sample
($N_{ALL} = 4782$), of all star-forming galaxies (SF+SSB;
$N_{TSF} = 2750$), and of passive galaxies ($N_{P} = 2032$),
as a function of the local galaxy density $\rho$, 
is shown in Fig. \ref{9}. Passive galaxies have a
distribution, relative to the curve for all galaxies, skewed towards
high density values, whereas the opposite tendency
is seen for star-forming galaxies. This indicates that there is, indeed, 
a star formation--environment relation, in the sense that the population
of star-forming galaxies decreases in denser environments.

\begin{figure}
\centerline{\includegraphics[width=84mm]{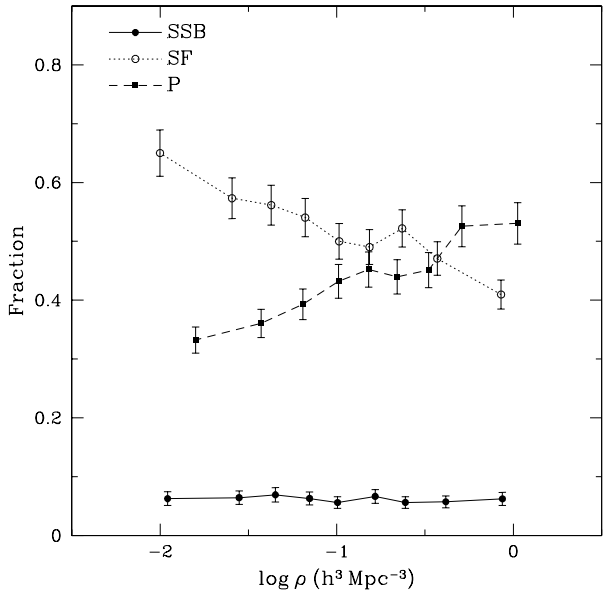}}
\caption{\small Fraction of galaxies in each spectral class 
as a function of the local galaxy density. 
Each density bin contains about the same number of galaxies. 
The error bars are computed assuming a Poisson statistics.}
\label{10}
\end{figure}

Fig. \ref{10} presents the relation between the fractions of galaxies in 
different spectral classes -- SSB, SF and P -- as a function of $\rho$.
Each density bin contains about the same number of galaxies 
(ranging from 32, for SSB, to 274, for the SF class);
the errors were computed assuming a  Poisson statistics. 
The figure shows that the fraction of passive galaxies increases with 
increasing density; in contrast, the SF fraction decreases with 
$\rho$. This figure also shows that the incidence of SSB is 
essentially environment independent: their fraction is $\sim 6$ per 
cent everywhere. The behaviour of the populations with local density 
is smooth and any threshold is seen over this range of densities.

Assuming that the trends shown in Fig. \ref{10} are
linear, we may examine their significance through the Pearson correlation
coefficient, $r$ \citep[e.g., ][]{press}.  The P and SF classes have $r$ 
equal to $0.97$ and $-0.96$, respectively, and the probability $p$
of the null hypothesis of zero correlation is equal to $1.7 \times 10^{-5}$ 
and $4.0 \times 10^{-5}$, for 
P and SF. For the SSB we have $r=-0.39$ and $p=0.30$, indicating that 
there is not a significant correlation between fraction and
the local density for this class.
These results are robust, even for the less populated SSB class, as we 
verified by repeating this analysis considering only galaxies with
errors in EW([{\sc O~ii}]) and EW(H$\delta$) below the median. 
All the trends have been confirmed, at similar levels of 
significance.

Since the fraction of star-forming galaxies decreases with
increasing density whereas that of SSBs is relatively density 
insensitive, in dense environments SSBs become relatively more frequent 
among star-forming galaxies. 
This behaviour had been noticed already in a sample of galaxies in the 
Shapley supercluster \citep{cuevas}. From a study of 8 Abell 
clusters, \citet{moss00} have also concluded that the fraction of 
SSBs among spirals increases from regions of lower to higher local 
galaxy surface density.

\begin{figure}
\centerline{\includegraphics[width=84mm]{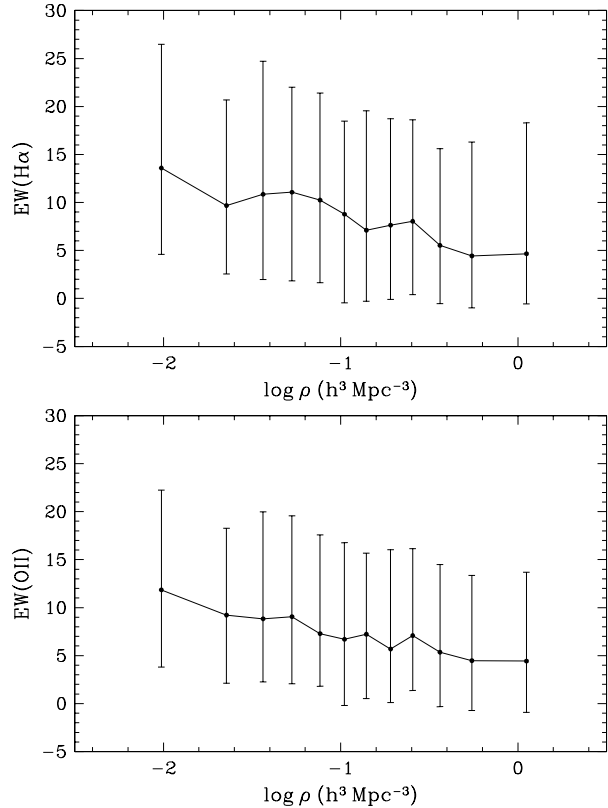}}
\caption{\small H$\alpha$ and [{\sc O~ii}] equivalent width distributions
as a function of local density of galaxies. The filled circles connected by
a solid line represent
the median values of each distribution and the error bars 
are the respective quartiles. 
Each density bin contains about the same number of galaxies (399).}
\label{11}
\end{figure}

It is also interesting to examine in this sample
the environmental dependence of the EWs 
of emission lines that are sensitive to star formation.
In Fig. \ref{11} we show the distributions of EW(H$\alpha$)
and EW([{\sc O~ii}]) as a function of local galaxy density, for the
$4782$ objects in our sample. The filled circles linked by solid lines 
represent the median values for each density bin (containing $\sim 399$ galaxies),
whereas the error bars are the respective quartiles. 
The EWs of H$\alpha$ and [{\sc O~ii}] tend to decrease regularly
with increasing density. The decline of these EWs with $\rho$ is 
present in the upper and lower quartiles, as well as in the median 
values. Assuming again linear correlations, we obtain in this case
$r$ equal to $-0.94$ and $-0.96$, and $p$ equal to $6.2 \times 10^{-6}$ 
and $1.2 \times 10^{-6}$, for EW(H$\alpha$)
and EW([{\sc O~ii}]), respectively. Thus, we may conclude the mean star 
formation properties of field galaxies, probed by these EWs, varies 
smoothly with the environment, even at low densities.

\subsection{Aperture bias}

At this point it is convenient to revisit the problem of aperture bias 
(see \ref{spectral}). This effect was not considered in the sample 
selection nor in the previous analysis. However, it
may introduce a redshift dependence in the measured
galaxy spectra, since the fraction of galaxy light received by a fiber 
increases with increasing distance. \citet{zaritsky95}, based on an
analysis of LCRS spectra, suggest that for reshifts $z > 0.05$ 
the effect is minimized. This redshift is exactly the upper limit 
of our sample and, thus, it is necessary to
verify whether our results are significantly affected by this bias.

\begin{figure}    
\centerline{\includegraphics[width=84mm]{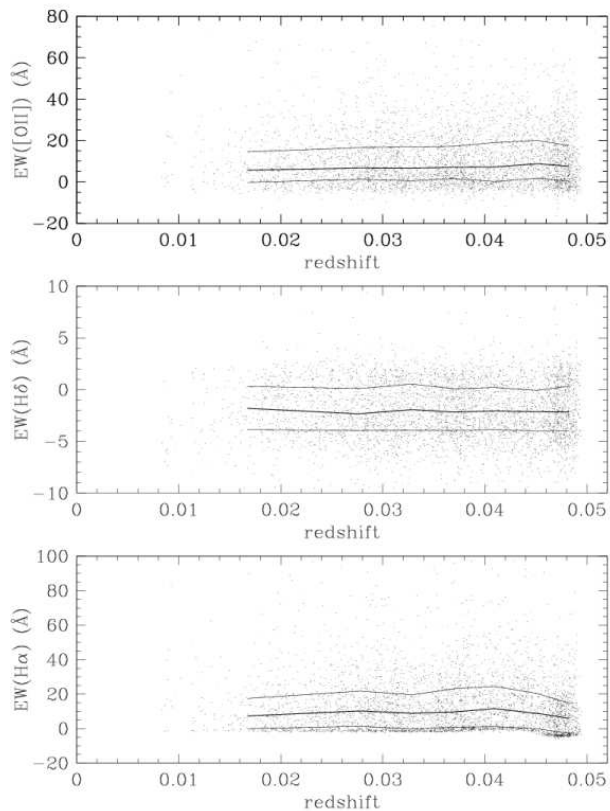}}
\caption{\small Distribution of EWs as a function of the redshift:
a) EW([{\sc O~ii}]), b) EW(H$\delta$), c)EW(H$\alpha$).
It is also shown, for each distribution, their median values and the 
respective quartiles (as solid lines).}
\label{12}
\end{figure}

For that, we investigated the behaviour of the EWs of [{\sc O~ii}], 
H$\delta$, and H$\alpha$ as a function of redshift. 
We divided the sample in several redshift bins containing 
the same number of objects, and computed the median value and the
quartiles of each spectral indice in each bin. 
Fig. \ref{12} shows as solid lines the median values of each 
distribution, as well as their respective quartiles. Each redshift
bin contains the same number of objects. Clearly, 
the median values of the equivalent widths of the three lines
analysed here do not show any trend with the redshift, meaning that
our measurements seem to be unaffected by the aperture bias.

\section{Discussion}\label{discussion}

The environmental behaviour of the fractions of
SF and P spectral classes, shown in
Fig. \ref{10}, complements results previously obtained by other authors
mainly for the SFR.
For example, \citet{ba98,ba99} found that cluster galaxies have lower 
star-formation rates than field galaxies with similar disk-to-bulge ratio
and luminosity. \citet{hashi98} used the Las Campanas Redshift Survey
data to study the relation between local density and star formation in 
galaxies with same concentration index. Their results show that star-forming
galaxies are preferentially found in low-density environments, and the
authors point out that the SFR of galaxies with similar structure is
sensitive to the local galaxy density. Similar results were also
obtained by \citet{carter}, using the spectroscopic data from
the 15R-North galaxy redshift survey. Here we have considered a
sample of field galaxies, showing that the fractions of the P and SF 
spectral classes vary smoothly with the local density.

\citet{lewis02} have used 2dFGRS spectra to investigate
the star formation rate of galaxies in different environments around clusters
using a parameter $\mu^*$ which is proportional to EW(H$\alpha$).
They found that there is, indeed, a correlation between SFR and local
projected density which is significant only for projected densities 
above $\sim$ 1 galaxy Mpc$^{-2}$ (for galaxies brighter than $M_b=-19$,
assuming $H_0 = 70$ km s$^{-1}$ Mpc$^{-1}$), that corresponds 
approximately to the mean density at the cluster virial radius.
Using data from the first release of the SDSS project
(Early Data Release, EDR), \citet{gomez02} also obtain a
relation between SFR and projected density. They show that there is a 
characteristic break in the SFR distribution at a local projected galaxy density of 
$\sim 1 h^{-2}_{75}$Mpc$^{-2}$, corresponding to a clustercentric
radius of $\sim$ 3--4 virial radii. Below this density the SFR increases
only slightly, whereas at denser regions it is strongly suppressed.
Thus, if one assumes that the equivalent width of H$\alpha$, 
for instance, is proportional to the star formation rate of a galaxy,
like assumed by \citet{lewis02}, our results are similar to those
obtained by \citet{gomez02} for regions farther than 3 virial radii
of cluster cores. Thus, the correlation between the H$\alpha$ and 
[{\sc O~ii}] equivalent widths and the local density shown in Fig. \ref{11} 
reinforce the idea that even low-density environments play an important role 
in stablishing the fraction of passive and/or star forming galaxies.

Our results also show that the fraction
of star-forming galaxies varies smoothly along all 
range of densities covered by our sample, suggesting that the population
of galaxies showing evidences of star formation activity seems to be 
affected by the environment everywhere. This is an important issue that
needs to be clarified, because it may provide important hints about the
physical mechanisms that are behind the relation between star formation
and galaxy environment. Indeed, if only environments with relatively
large densities affects star formation, processes that occur mainly in
clusters, like ram-pressure stripping or galaxy harassment may be
considered the drivers of the trends observed between star formation and
environment; on the other hand, if this trend is present for galaxies in 
clusters as well as for those in the field, as our results support,
mechanisms that act everywhere, like tidal interactions, should be 
playing a relevant role.

At this point it is convenient to discuss the behaviour of SSB galaxies.
We have shown in Section 3 that the fraction of these galaxies is
essentially independent of their luminosity and local number density, 
at least for the range of values of these quantities present in our sample.
We have also shown that an increase in density tends 
to decrease the fraction of star-forming galaxies, and then the observed
independence of the SSB fraction with $\rho$ is somewhat unexpected.
In fact, such a behaviour may be an indication that interactions,
either between galaxies, or between a galaxy and its environment, may trigger
a starburst in a galaxy.
\citet{moss00} suggest that tidal effects due to gravitational 
interactions (either galaxy$-$galaxy, galaxy$-$group or galaxy$-$cluster) 
originate bursts of star formation and morphological distortions 
in spiral galaxies, with a  frequency increasing within high-density
environments. Similar results have been reported by \citet{hashi00}. 
\citet{davis97}, in a study of poor groups of galaxies, also suggest
that the interaction of a galaxy with their neighbours increases 
its star formation rate; however, they point out that morphological 
peculiarities produced by tidal effects may leave the {\sc H~i} disks 
more vulnerable to external hydrodynamic forces, turning subsequent gas 
removal more efficient. Consequently, any star-formation burst in high 
density environments should be short-lived in this scenario.

In order to verify whether tidal interactions are actually inducing
the starbursts that characterize SSB galaxies, we tried to verify
whether these galaxies in our sample do show any evidence of
morphological distortions. For this, we have examined their images,
using the SuperCOSMOS Sky
Survey\footnote{\texttt{http://www-wfau.roe.ac.uk/sss/}}\citep{hambly}.
We divided the SSB galaxy sample in two groups containing 50
objects in each one, comprising galaxies located in the bin of lowest
galaxy density ($-2.44  < \log \rho$ (Mpc$^{-3}$ $h^{-3}$)$\,< -1.52$)
and in the bin of highest density
($-0.40  < \log \rho$ (Mpc$^{-3}$ $h^{-3}$)$\,< 0.46$),
respectively. In low-density regions, the fraction of SSBs showing any
kind of peculiarity is $\sim 10$ per cent. When we analyse images of the
galaxies in the densest environments, this fraction increases to
$\sim 30$ per cent. As a conclusion, although tidal
interactions may not be the only mechanism producing the starburst
that characterizes SSB galaxies, they are probably having a major
role in inducing the large rates of star-formation observed in
these objects, even in the low-density regions of the field.

\section{SUMMARY AND CONCLUSIONS}
 
In this paper we have investigated the environmental dependence of 
the population of star-forming galaxies with a volume-limited sample of field galaxies
extracted from  the final release of the 2dF Redshift Galaxy Survey 
and containing $4782$ objects. 
We have adopted a spectral classification that 
characterizes the star formation through the equivalent widths of
[{\sc O~ii}]$\lambda$3727 and H$\delta$. The environment is described by the
local spatial density of galaxies. We have then analysed the 
distribution of fractions for each spectral class as a function of
density, mainly, and luminosity. Our main findings are the following:

\begin{enumerate}

\item The star-forming galaxies in our sample
are mostly low-luminosity objects,
whereas brighter objects include mainly passive galaxies that do not 
present significant activity of star formation. The same behaviour is
noted when we analyse the luminosity dependence of star formation
indicators like H$\alpha$ and
[{\sc O~ii}] equivalent widths: these quantities are higher for 
low-luminosity objects.
The short starburst galaxies (SSB) do not show strong
trends with luminosity, except for magnitudes at the faint end of our sample,
where their fraction seems to increase.

\item The environmental distribution of spectral classes indicates
that star-forming galaxies (SF) and passive galaxies (P) display a 
distinct behaviour relative to their habitat: the P fraction 
increases with increasing density; in contrast, the SF 
fraction decreases with $\rho$. On the other hand, the fraction of 
SSB galaxies is essentially environment independent, and relative 
to all star-forming galaxies, the fraction of SSBs increases
with increasing density.

\item The smooth relation between
the fraction of star-forming galaxies and local density
indicates that process affecting 
star formation activity do act everywhere, suggesting that mechanisms
like tidal interactions -- that act in all environments -- 
may play a relevant role on the star formation properties of galaxies.

\end{enumerate}

In the hierarchical scenario of galaxy formation, as galaxy clustering
evolves, the density around a galaxy tends to increase, in all
environments. Higher density probably means more interactions. Our
results give support to this broad view, because the fraction of
star-forming galaxies seems to
be affected by the environment in all range of densities covered by our 
sample. Thus, there are physical 
mechanisms that act efficiently on the star formation in 
all environments, not only in the high-density regions of galaxy 
clusters. 

\section*{Acknowledgments}

We thank  the 2dFGRS Team for making available the data 
analysed here. We also thank Peder Norberg and Shaun Cole, who provided
the 2dFGRS mask software used in the simulations. We also thank 
T. Goto and M. Balogh for very interesting comments on a previous version of
this paper, as well as an anonymous referee for helpful 
comments and suggestions. This work has benefitted 
from financial support from Fapesp, CNPq, and CCINT.

\bsp

\label{lastpage}

\end{document}